\pdfoutput=1
\documentclass[final,authoryear,5p]{elsarticle}
\usepackage{natbib}
\usepackage{epsfig}
\usepackage{textcomp}
\usepackage{amssymb}
\usepackage{amsmath}
\usepackage{url}
\usepackage{fixltx2e} 
\usepackage{booktabs}
\journal{Advances in Space Research}
\begin{document}
\begin{frontmatter}
\title{Orbital Debris-Debris Collision Avoidance}
\author{James Mason\corref{cor}}
\address{NASA Ames Research Center and Universities Space Research Association, Moffett Field, MS202-3, CA 94035, USA}
\cortext[cor]{Corresponding author}
\ead{james.mason@nasa.gov}
\author{Jan Stupl}
\address{Center for International Security and Cooperation, Stanford University, 616 Serra Street, CA 94305, USA}
\author{William Marshall}
\address{NASA Ames Research Center and Universities Space Research Association, Moffett Field, MS202-3, CA 94035, USA}
\author{Creon Levit}
\address{NASA Ames Research Center, Moffett Field, MS202-3, CA 94035, USA}
\begin{abstract}
We focus on preventing collisions between debris and debris, for which there is no current, effective mitigation strategy. We investigate the feasibility of using a medium-powered (5\,kW) ground-based laser combined with a ground-based telescope to prevent collisions between debris objects in low-Earth orbit (LEO). The scheme utilizes photon pressure alone as a means to perturb the orbit of a debris object. Applied over multiple engagements, this alters the debris orbit sufficiently to reduce the risk of an upcoming conjunction. We employ standard assumptions for atmospheric conditions and the resulting beam propagation. Using case studies designed to represent the properties (e.g. area and mass) of the current debris population, we show that one could significantly reduce the risk of nearly half of all catastrophic collisions involving debris using only one such laser/telescope facility. We speculate on whether this could mitigate the debris fragmentation rate such that it falls below the natural debris re-entry rate due to atmospheric drag, and thus whether continuous long-term operation could entirely mitigate the Kessler syndrome in LEO, without need for relatively expensive active debris removal.
\end{abstract}
\end{frontmatter}
\parindent=0.5 cm
\section{Introduction}
The threat of catastrophic or debilitating collisions between active spacecraft and orbital debris is gaining increased attention as prescient predictions of population evolution are confirmed. Early satellite environment distribution models showed the potential for a runaway ``Kess\-ler syndrome'' of cascading collisions, where the rate of debris creation through debris-debris collisions would exceed the ambient decay rate and would lead to the formation of debris belts \citep{Kessler1978}. Recorded collisions events (including the January 2009 Iridium 33/Cosmos 2251 collision) and additional environmental modeling have reaffirmed the instability in the LEO debris population. The latter has found that the Kess\-ler syndrome is probably already in effect in certain orbits, even when the models use the extremely conservative assumption of no new launches \citep{Liou2008, Liou2009}.

In addition to the UN COPUOS's debris mitigation guidelines, collision avoidance (COLA) and active debris removal (ADR) have been presented as necessary steps to curb the runaway growth of debris in the most congested orbital regimes such as low-Earth sun synchronous orbit \citep{Liou2009}. While active spacecraft COLA does provide some reduction in the growth of debris, alone it is insufficient to offset the debris-debris collisions growth component \citep{Liou2011}. \citet{Liou2009} have suggested that stabilizing the LEO environment at current levels would require the ongoing removal of at least 5 large debris objects per year going forward (in addition to a 90$\%$ implementation of the post mission disposal guidelines). Mission concepts for the removal of large objects such as rocket bodies traditionally involve rendezvous, capture and de-orbit. These missions are inherently complex and to de-orbit debris typically requires $\Delta v$ impulses of order 100\,m/sec, making them expensive to develop and fly. Additionally, a purely market-based program to solve this problem seems unlikely to be forthcoming; many satellite owner/operators are primarily concerned with the near term risk to their own spacecraft and not with long term trends that might endanger their operating environment, making this a classic ``tragedy of the commons'' \citep{Hardin1968}. The cost/benefit trade-off for active removal missions makes them unlikely to be pursued by commercial space operators until the collision risk drives insurance premiums sufficiently high to warrant the investment.

To quantify this risk one can look to an example: ESA routinely performs detailed conjunction analysis on their ERS-2 and Envisat remote sensing satellites \citep{Klinkrad2005}. Although the number of conjunctions predicted annually for Envisat by ESA's daily bulletins is in the hundreds, only four events had very high collision probabilities (above 1 in 1,000). None of these conjunctions required avoidance maneuvers after follow-up tracking campaigns reduced orbital covariances, or uncertainties \citep{Klinkrad2009}. While several maneuvers have been required since then, the operational risk is still insufficient to provide incentive for large scale debris remediation effort and this highlights the need for low-cost, technologically mature, solutions to mitigate the growth of the debris population and specifically to mitigate debris-debris collisions which owner/operators can not influence with collision avoidance. Governments remain the key actors needed to prevent this tragedy of the commons that threatens the use of space by all actors.

Project ORION proposed ablation using ground-based lasers to de-orbit debris \citep{Campbell1996}. This approach requires MW-class continuous wave lasers or high energy pulses (of order 20\,kJ per 40ns pulse) to vaporize the debris surface material (typically aluminum) and provide sufficient recoil to de-orbit the object. ORION showed that a 20\,kW, 530\,nm, 1\,Hz, 40\,ns pulsed laser and 5\,m fast slewing telescope was required to impart the $\Delta v$ of 100-150\,m/sec needed to de-orbit debris objects. This was technically challenging and prohibitively expensive at that time \citep{Phipps1996}. Space-based lasers have also been considered, but ground-based laser systems have the advantage of greatly simplified operations, maintenance and overall system cost.

In this paper we propose a laser system using only photon momentum transfer for debris-debris collision avoidance. Using photon pressure as propulsion goes back to the first detailed technical study of the solar sail concept \citep{Garwin1958}. The use of lasers to do photon pressure propulsion was first proposed by \citet{Forward1962}. For the application of this to collision avoidance, a $\Delta v$ of 1\,cm/s, applied in the anti-velocity direction results in a displacement of ~2.5\,km/day for a debris object in LEO. This along track velocity is far larger than the typical error growth of the known orbits of debris objects. Such small impulses can feasibly be imparted only through photon momentum transfer, greatly reducing the required power and complexity of a ground based laser system. Additionally, this reduces the potential for the laser system to accidentally damage active satellites or to be perceived as a weapon.

\citet{Levit2010} provide details of ongoing conjunction analysis research at NASA Ames Research Center, including all-on-all conjunction analysis for the publically available U.S Strategic Command (USSTRATCOM) Two Line Element (TLE) catalog and simulated future catalogs of up to 3 million objects on the Pleiades supercomputer. Their paper also presents early results suggesting that a high accuracy catalog comparable to the USSTRATCOM special perturbations (SP) catalog can be generated from the publicly available TLEs; sufficiently accurate to allow collision avoidance with $\Delta v$ in the sub-cm/s range.

This laser COLA scheme was first proposed in \citet{Levit2010} and it is the purpose of this paper to give a more detailed analysis. We focus on assessing the effectiveness of a laser facility for making orbit modifications. The system proposed in this paper uses a 5-10\,kW continuous wave laser mounted on a fast slewing 1.5\,m optical telescope with adaptive optics and a sodium guide star, which allows the laser beam to be continuously focused and directed onto the target throughout its pass.

We start by discussing the underlying physical phenomena, then describe the baseline system and the design of our case study. We conclude by presenting the results of a case study, summarizing the potential applications and identifying further research.

\section{Methodology: Perturbing LEO debris orbits with Radiation Pressure}
\begin{figure}
\begin{center}
\includegraphics*[width=8.8cm,angle=0]{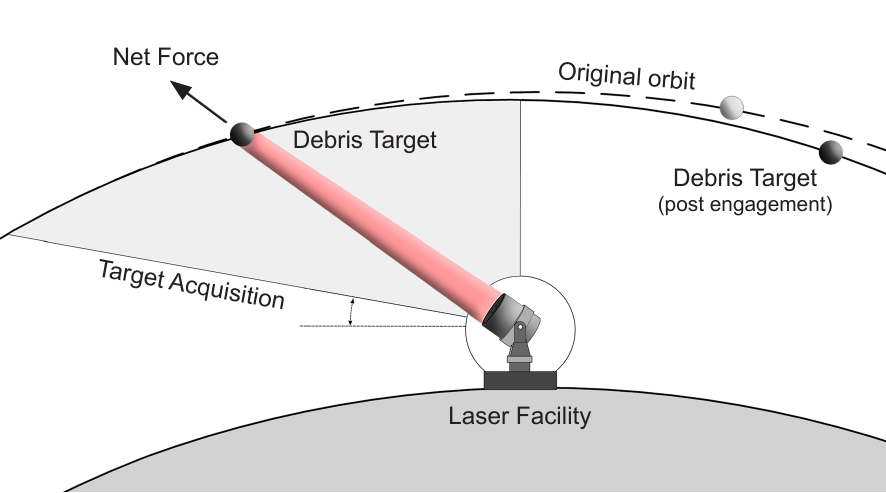}
\end{center}
\caption{Schematic of laser system and operations}
\label{figure1}
\end{figure}

In order to assess the feasibility of a collision avoidance scheme based on laser applied radiation pressure, we simulate the resulting orbit perturbations for a number of case studies. The laser radiation adds an additional force to the equations of motion of the irradiated piece of debris, which are then evaluated by a standard high precision orbital propagator. Application of a small $\Delta v$ in the along-track direction changes the orbit's specific energy, thus lowering or raising its semi-major axis and changing its period (illustrated in Fig \ref{figure1}). This allows a debris object to be re-phased in its orbit, allowing rapid along-track displacements to grow over time.

Comprehensive all-on-all conjunction analysis would i\-den\-ti\-fy potential debris-debris collisions and prioritize them according to collision probability and environmental impact (a function of object mass, material, orbit, etc.), as well as screening out conjunctions for which the facility is unable to effect significantly (e.g. one involving two very massive or two very low $A/M$ debris objects). For conjunctions with collision probabilities above a certain ``high risk'' threshold (say 1 in 10,000) we would then have the option of choosing the more appropriate object (typically lower mass, higher $A/M$) as the illumination target. Objects of lower mass will be perturbed more for a given force per unit area. Below we discuss how to approximate the area to mass ratio of the object and how to model the displacement that is possible with a given system.

\subsection{Assessing radiation pressure}
Radiation pressure is a result of the photon momentum. If a piece of debris absorbs or reflects incoming photons, the momentum transferred leads to a small, but significant force. As described in the literature \citep{McInnes1999}, the resulting force per unit area, i.e. the radiation pressure, is
\begin{equation}\label{radpressbasic}
F/A = C_r \times p = C_r \times I/c
\end{equation}
where $A$ is the illuminated cross section, $I$ is the intensity of the radiation, $C_r$ is the radiation pressure coefficient of the object and $c$ is the speed of light. $C_r$ can take a value from 0 to 2, where $C_r = 0$ means the object is translucent and $C_r = 2$ means that all of the photons are reflected (i.e. a flat mirror facing the beam). An object which absorbs all of the incident photons (i.e. is a black body) has $C_r = 1$. For constant intensities, the resulting force can be obtained by simple multiplication. However, for larger pieces of debris, the intensity will vary over the illuminated cross section. Hence, we choose to implement a more accurate description for our simulation, integrating over the illuminated cross-section.
\begin{equation}\label{radforcebasic}
F = C_r/c \int I(x,y) \ dA
\end{equation}

The intensity distribution $I(x,y)$ at the piece of debris depends on the employed laser, its output power and optics, and the atmospheric conditions between the laser facility and the targeted piece of debris. In the simplest case, $I(x,y)$ will be axisymmetric $I=I(r)$ and follow a Gaussian distribution \citep{Siegman1986}
\begin{equation}\label{gaussintensity}
I(L,r) = I_{0} e^{-2r^{2}/w(L)^2}
\end{equation}
where $I_{0}$ is the maximum intensity of the beam and $w$ is the beam width, defined as the radius where the intensity drops to $1/e^{2}$ of the maximum $I_0$ in a given plane at a distance $L$ from the laser. $I_{0}$ depends on the beam width, as a larger beam width will lead to the energy being distributed over a larger area. The beam width is a function of the distance $L$ between the laser and the debris. $w$ is somewhat controllable but depending on the laser, its optics and atmospheric conditions, there is a lower limit for the beam width.

The lower limit for $w_{0(min)}$ for an ideal laser propagating in a vacuum is given by the diffraction limit,
\begin{equation}\label{difflimitspot}
w_{0(min)} \approx \lambda L/D
\end{equation}
where $\lambda$ is the wavelength of the laser, $D$ is the diameter of the focusing optic and $L$ is the distance between the optic and the piece of debris \citep[p. 676]{Siegman1986}.

Assuming an object in a 800\,km orbit, passing directly overhead a station which uses a solid state laser with a wavelength of 1\,\textmu m and a focusing optic with a 1.5\,m diameter, a minimum beam width of 0.6\,m would result. Increasing the beam width is always possible, but in order to maximize the force applied, we assume the beam width is at a lower limit.

In the case of a real laser facility the atmosphere has two major effects on beam propagation. First, different constituents will absorb and/or scatter a certain amount of energy. Second, atmospheric turbulence leads to local changes in the index of refraction, which increases the beam width significantly. In addition, the resulting time-dependent intensity distributions might not resemble a Gaussian at all. However, laser engagements in our case will take place over time frames of minutes so we adopt a time-averaged approach. As common in this field, we choose an extended Gaussian model, where the minimum beam width is increased by a beam propagation factor, leading to a reduced maximum intensity. It has been shown that this ``embedded Gaussian'' approach is valid for all relevant intensity distributions, allowing simplified calculations \citep{Siegman1991}. Even if the Gaussian model might not resemble the actual intensity distribution, the approach ensures that the incoming time-averaged total intensity is correct \citep{ Siegman1998}. The resulting intensity at a distance $L$ from the laser depends on the conditions on a given path $\vec{L}$ through the atmosphere.

\begin{align}\label{intensity}
I(\vec{L},r) = &S_{sum}(\vec{L})\times \tau(\vec{L})\nonumber \times \frac{2P}{\pi w_{0(min)}^2} \\ &\times exp\left(-2 S_{sum}(\vec{L})\times \frac{r^2}{ w_{0(min)} ^2}\right)
\end{align}
where $P$ is the output power of the laser and $w_{0(min)}$ is the minimum beam diameter in a distance $L$ calculated according to equation \ref{difflimitspot}. This lower limit is increased by the Strehl factor $S_{sum}$. The total transmitted power is reduced by a factor $\tau$, accounting for losses through scattering and absorption. $\tau$ and $S_{sum}$ depend on the atmospheric path and this path changes during the engagement as the debris crosses the sky. $\tau$ and $S_{sum}$ are calculated for each time step by integrating atmospheric conditions along the path at that time. We use the standard atmospheric physics tool MODTRAN 4 (Anderson, 2000) to calculate $\tau$. $S_{sum}$ is a cumulative factor that includes the effects of a less than ideal laser system and optics in addition to turbulence effects. To assess turbulence effects we use the Rytov approximation. The Rytov approximation is a statistical approach commonly used in atmospheric optics that combines a statistical turbulence model and perturbation theory to modify the index of refraction in the wave equation. The theoretical background and details of our numerical approach are described elsewhere \citep[appendix A]{Stupl2010}, \citep[chapter 2]{Stupl2008}, including additional references therein on atmospheric optics and turbulence.

Our calculations show that turbulence reduces the effectiveness of the system by an order of magnitude - principally by increasing the effective divergence. To counter those effects, we assume that an adaptive optics system with an artificial guide star is used. Such a system measures the effects of turbulence and counters them using piezoelectric deformable mirrors. The correction has to be applied in real time, as local turbulence changes rapidly and the guide star moves across the sky as the telescope tracks the target. Adaptive optics performance varies depending on the degree of turbulence in the path of the beam and the technical capabilities of the adaptive optics system.

Physical properties of space debris objects vary and for a majority of objects some parameters are unknown. This makes accurate modeling difficult. A discussion of the key parameters and our assumptions follows.

\subsection{Area to Mass ratio}
\begin{figure}
\begin{center}
\includegraphics*[width=8.8cm,angle=0]{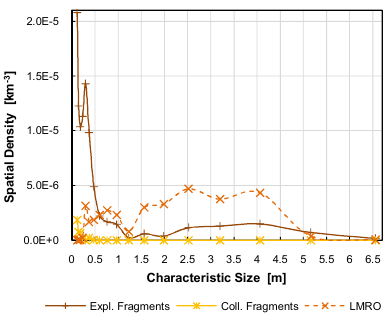}
\end{center}
\caption{MASTER2005 spatial density in sun synchronous orbit between 600\,km and 1100\,km altitude. Note that Launch and Mission Related Objects (LMRO) include active, maneuverable satellites. Additionally, this figure does not include the Fengyun IC and Iridium/Cosmos debris (we are awaiting new MASTER data)}
\label{spatialdens}
\end{figure}

The acceleration from photon pressure on a debris target is proportional to the object's area and inversely proportional to its mass. To accurately model the photon pressure from a beam of width $w$ on an object, both area and mass need to be independently known. Since this research presents an initial feasibility investigation, the dimensions for a random set of debris objects can be inferred from statistical data on debris size. The ESA Master model provides statistics on observed characteristic size distributions (shown in Fig \ref{spatialdens}) for objects in our region of interest, namely sun-synchronous LEO - the most problematic region for debris-debris collisional fragmentation \citep{Oswald2006}.

Launch and Mission Related Objects, including rocket upper stages and intact satellites, greatly dominate the total mass of objects in LEO and are generally too massive to be effectively perturbed using photon pressure alone. The implication is that this scheme, as presented, would likely be ineffective at preventing collisions between two massive objects such as rocket bodies or intact spacecraft. However, over 80\% of all catalogued objects in sun-synchronous LEO are debris resulting from explosions or collisions, and a significant proportion of these may be effectively perturbed using photon pressure alone since fragments typically have high A/M ratios and low masses \citep{Anselmo2010}. The efficacy of the laser photon pressure approach as a long term debris remediation tool therefore depends on the proportion of collisions that involve high area to mass ratio objects in general, and debris fragments in particular.The ballistic drag coefficient, defined as the product of the dimensionless drag coefficient $C_d$ and the area to mass ratio $A/M$, for an object is given by \citep{Vallado2007}:
\begin{equation}\label{ballisticdrag}
B = C_d \times A / m = 12.741621 B^\star
\end{equation}
where $B^\star$ (BSTAR) is a free parameter of the orbit determination process used to generate TLEs. This relationship holds for an atmospheric model that does not vary with solar activity but in the case of low solar activity Eq. \ref{ballisticdrag} systematically underestimates the ballistic coefficient for debris fragments, sometimes by multiple orders of magnitude \citep{Pardini2009}. Additionally, the difficulties in tracking irregular and small debris objects suggests that $B^\star$ for debris objects is less accurate than for large rocket bodies or satellites. In fact, a number of objects were found in the catalog with no $B^\star$ information at all. 

A more accurate method for determining the ballistic coefficient is to rescale $B$ by fitting the observed decay of the semi-major axis of the object over a long period, using an accurate atmospheric model and a high accuracy orbit integrator \citep{Pardini2009}. We implemented this method by downloading 120 days of TLEs for each debris object and then using a standard high precision orbit propagator to fit the ballistic coefficient to the observed decay of semi-major axis. Assuming $C_d = 2.2$, a reasonable value for the $A/M$ ratio of an object can be estimated.

\subsection{Spin state and Reflectivity}
The spin state of a debris object introduces a degree of randomness into calculating the response to directed photon pressure. The momentum transferred from absorbed photons will be in the incident beam direction. For a tumbling target the force vector due to reflection will be varying during the engagement, since there will be a component of the force orthogonal to the laser incidence vector, and for most targets the laser will also induce a torque about the center of mass, which we ignore for the present. We follow the ORION study and assume that collision and debris fragments above 600\,km will be rapidly spinning \citep{Phipps1996}. On average, for quickly tumbling objects, orthogonal force vectors (due to specular reflection) will be zero and the net force vector due to diffuse reflection will be directed parallel to the laser beam. 

Mulrooney and Matney (2007) suggest that debris has global albedo value of 0.13 which in the general case would give $C_r = 1.13$. However, we make a conservative assessment and neglect the effects of diffuse reflection, assuming a force parallel to the laser beam according to equation \ref{radforcebasic}, where $C_r = 1.0$. In reality, the resulting net force will likely be larger and for slowly spinning objects the net force will not be in the beam direction. 

In an operational setting, one would propagate forward a range of laser vector $\Delta v$ (associated with unknowns in $C_r$, $A/M$ etc.) and a range of orthogonal $\Delta v$ to account for uncertainties in object forms and spin states. The implications of the engagement could then be assessed using the resulting error ellipsoid of the maneuver e.g. to ensure that the maneuver would not cause future conjunctions with other objects in the debris field. For the purpose of this study we also assume that the illuminated cross sectional area is equal to the effective average cross section, as determined by our long term estimation of the drag area. This is equivalent to approximating the rapidly tumbling object as a sphere of radius equal to this average drag area.

\subsection{Implementation}
For determining the ballistic coefficient of an object from the decay of its semi-major axis we used AGI's Satellite Tool Kit (STK) and an iterative differential corrector to fit a high precision orbit to the object's historical TLEs.

We developed a model for laser propagation in an atmosphere as per Section 2.1 using MATLAB and MODTRAN 4. Target objects were propagated using a high precision propagator in STK, accounting for higher-order gravitational terms, a Jacchia-Roberts atmospheric model, observed solar flux and spherical solar radiation pressure. Laser engagements were modeled by utilizing the MATLAB-STK scripting environment, allowing the evaluation of the laser intensity and resulting photon pressure at each time step.

\section{Baseline System}
Past studies have looked into active debris removal using laser ablation. While these favorably assessed the feasibility of the approach, none of those systems have been developed and tested. One reason for this is their reliance on what are traditionally military-class systems. These are generally not commercially available or are one-of-a-kind experimental systems, making them very expensive and difficult to obtain. To avoid those shortfalls, we chose to restrict this study to medium power commercially available lasers and to shorten development times and reduce overall cost we also restrict this study to commercially available off-the-shelf technology for other parts of the system where possible. Below we outline an example system that might be developed today at reasonable cost and the following case studies aim to assess whether collision avoidance is still possible with such a system.

\subsection{Laser}
The intensity that can be delivered to the target (described by equations \ref{difflimitspot} and \ref{intensity}) is proportional to the laser power and inversely proportional to the wavelength. The beam quality describes how well the laser beam can be focused over long distances, critical for targeting small debris objects. Atmospheric transmittance and technical constraints puts restrictions on useful wavelengths. For targeting sun-synchronous objects the ideal laser facility location would be close to the poles and so the equipment should be low maintenance and ruggedized. Combining these requirements, and restricting our choice to lasers commercially available, we identified an IPG single mode fiber laser with a $1.06$\,\textmu m wavelength. It is electrically powered with no parts requiring alignment (or that can become misaligned) and is designed for 24/7 industrial applications. The beam quality of this laser is close to the diffraction limit ($M^2=1.2$) and the output power is adjustable up to 5\,kW (IPG, 2009).

IPG also manufacturers a 10\,kW version and better results can be obtained with this higher output power. This gives some latitude for the other parameters as doubling the output power is still possible, albeit at a higher cost. As an additional benefit, this low power (compared to military systems) makes the system's application as an anti-satellite weapon unlikely and thus avoids some of the potential negative space security implications.

\subsection{Beam Director and Tracking}
The laser is focused onto the debris using a reflecting beam director. The beam director will most likely be an astronomical class telescope, potentially modified to manage the thermal effects of continuous laser operation. Neglecting atmospheric effects the maximum intensity is proportional to this telescope's aperture. The beam director has to be rapidly slewed in order to track the debris and the required tracking tolerances become increasingly difficult to maintain as diameter and mass increase. Suitable 1.5\,m telescopes with fast slew capabilities are commercially available, for example from the company L3, and so we choose 1.5\,m as our baseline diameter.

Tracking accuracy for the L3 telescope is of order of $10^{-1}$ arc seconds, which may not be sufficient for tracking small debris in sun synchronous orbit \citep{L3COM2008}. Hence additional measures have to be taken. For laser satellite communications and directed energy applications, active tracking / closed-loop techniques have been developed which are able keep the target in the center of the view once it has been acquired (e.g. see \citet{Riker2007}). Acquisition is more difficult and satellite laser ranging techniques such as beam widening or search patterns will be needed to initially find the target. It will probably be necessary to use an imaging telescope coupled to the beam director to allow simultaneous guide star creation, beam illumination and target imaging for acquisition and tracking. The Mt. Stromlo facility operated by Electro Optic Systems (EOS) near Canberra, Australia is able to acquire and track debris of 5\,cm size up to 3000\,km range using a 100\,W average power pulsed laser and a $1.8$\,m fast slew beam director \citep{Smith2007}. This demonstrates that the target acquisition and tracking requirements can be met, although it may prove necessary to include a pulsed laser in the proposed system to allow for range filtering during target acquisition (as is done by SLR systems).

\subsection{Adaptive optics}
Restricting the laser system to a single 5-10\,kW facility means that sufficient laser intensities can only be reached if the effects of atmospheric turbulence are countered by adaptive optics. The effectiveness of such a system will depend on the turbulence encountered and the technical capabilities of the system.

In our calculations, we assume that the system’s capabilities for turbulence compensation are comparable to the system used in 1998 benchmark experiments \citep{Higgs1998, Billman1999}, which were conducted to test the proposed adaptive optics for the Airborne Laser missile defense project. The American Physical Society has compiled those results into a relationship of Strehl ratio vs. turbulence \citep[p. 323]{Barton2004} and we use this relationship in our numerical calculations to set the upper limit of the assumed adaptive optics performance. . While the ABL is a military system (and has much greater output power than necessary for COLA), the Large Binocular Telescope has shown a similar performance, reaching Strehl ratios up to $0.8$\citep{MPIA2010}. 

Turbulence effects must be measured in order to be compensated.. We assume that a laser guide star (positioned ahead of the target to account for light travel times) is used as a reference point source and compensation for de-focus and higher order turbulence effects is ideal. However, tip/tilt correction requires a signal from the real object and not the guide star. We calculate the negative impact of this so called tilt-anisoplanatism and lower the intensities accordingly. 

\subsection{Location and Atmospheric conditions}
The described system is designed to illuminate debris in sun-synchronous orbits, so to maximize engagement opportunities we favor a location as close as possible to the poles. Additionally, situating the facility at high altitude reduces the atmospheric beam losses and turbulence effects. An ideal site would be the PLATeau Observatory (PLATO) at Dome A in Antarctica, which is at 4\,km altitude and is in the driest region of the world. For comparison we also considered Maui and Mt. Stromlo, since they already have facilities that might be upgraded to test this concept, and a hilltop near Fairbanks, Alaska due to its high latitude and ease of access compared to arctic territory.

Atmospheric conditions will have a major impact on the performance of the system. Site selection and dome design will have to take this into account to minimize losses and down time. For this study we chose standard conditions for turbulence and atmospheric composition (Hufnagel-Valley 5/7 turbulence and the US Standard Atmosphere (1976), MODTRAN set to 365\,ppm CO$_2$, Spring/Summer conditions, and 23\,km surface meteorological range).

\subsection{Scalability}
While the laser parameters are readily available using a datasheet, tracking accuracy and adaptive optics performance are less certain. Since the effect of laser engagements is cumulative, one could both increase the power of the laser and use multiple stations, engaging debris from different locations, if adaptive optics performance or accurate tracking becomes more difficult than expected (or if one wants to do collision avoidance for lower $A/M$ or heavier debris objects). For example, by upgrading the laser to a 10\,kW model and having 3 or 4 facilities the effect of this system can be increased by an order of magnitude.

\subsection{Operational Considerations}
In general we want to lower the orbits of debris objects to reduce their lifetime so the optimal tasking of the laser-target engagement is to begin illuminating the target from the horizon and to cease the engagement when the target reaches its maximum elevation (simulated engagements start at 10$^\circ$ elevation to approximate acquisition delays). The main components of the net force for an overhead pass are in the anti-velocity and radial directions. Engaging during the full pass would result in a net radial $\Delta v$, which results in less rapid displacement over time from the original trajectory.

Target acquisition and tracking at the start of each engagement will produce track data and, if a pulsed laser is used for acquisition, ranging data similar to that produced by the EOS Space Debris Tracking System \citep{Smith2007}. This would allow orbit determination algorithms to reduce the error covariance associated with that object's orbit - helpful for space situational awareness (SSA) in addition to down range target re-acquisition. Laser campaigns would only need to continue until the collision risk has been reduced to an acceptable level - which can be either through improved covariance information and/or through actual orbit modification.

\section{Resulting Capabilities}
To quantify the effectiveness of this laser scheme on debris objects we start by demonstrating our method for an object of known mass and area. A discarded lens cap from the Japanese Akari IR space telescope was chosen as the demonstration object (U.S. Catalog ID: 29054). We nominally chose 01 January 2011 00:00:00 UTC as the starting time for all simulations. The lens cap is approximately a flattened hemispherical dome of mass 5\,kg, with a diameter of 80\,cm and a thickness of approximately 10\,cm. These parameters represent a large debris fragment. This lens cap orbits in a near circular orbit at about 700\,km altitude, with an inclination of $98.26^\circ$. 

\begin{figure}
\begin{center}
\includegraphics*[width=8.8cm,angle=0]{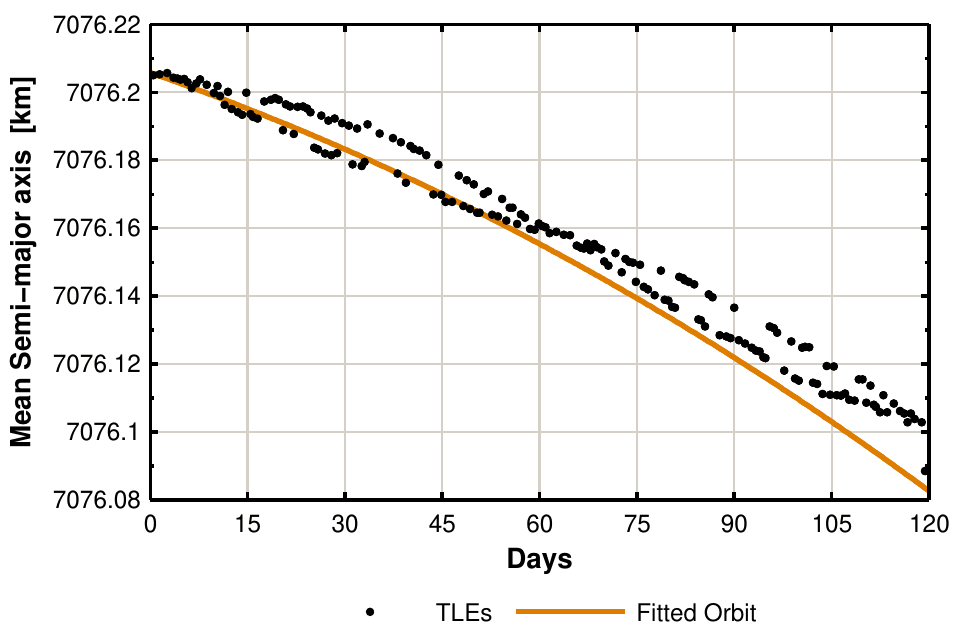}
\end{center}
\caption{Orbital decay of semi-major axis for Akari lens cap. The ``Fitted Orbit'' represents the orbit decay using the rescaled $A/M$ ratio, as fitted to the TLEs with a highly accurate special perturbations propagator.}
\label{capdecay}
\end{figure}

Fitting the observed orbital decay of the lens cap over 120 days (shown in Fig \ref{capdecay}) to derive the ballistic coefficient gave $A/M = 0.04$.
This is close to the minimum ballistic coefficient possible with the known object dimensions, suggesting that the lens cap has stabilized to present a minimum cross-section and to minimize drag forces. We initially use this area for radiation pressure calculations, even though the surface visible to the laser is likely to be larger.

\begin{figure*}
\begin{center}
\includegraphics*[width=14.4cm,angle=0]{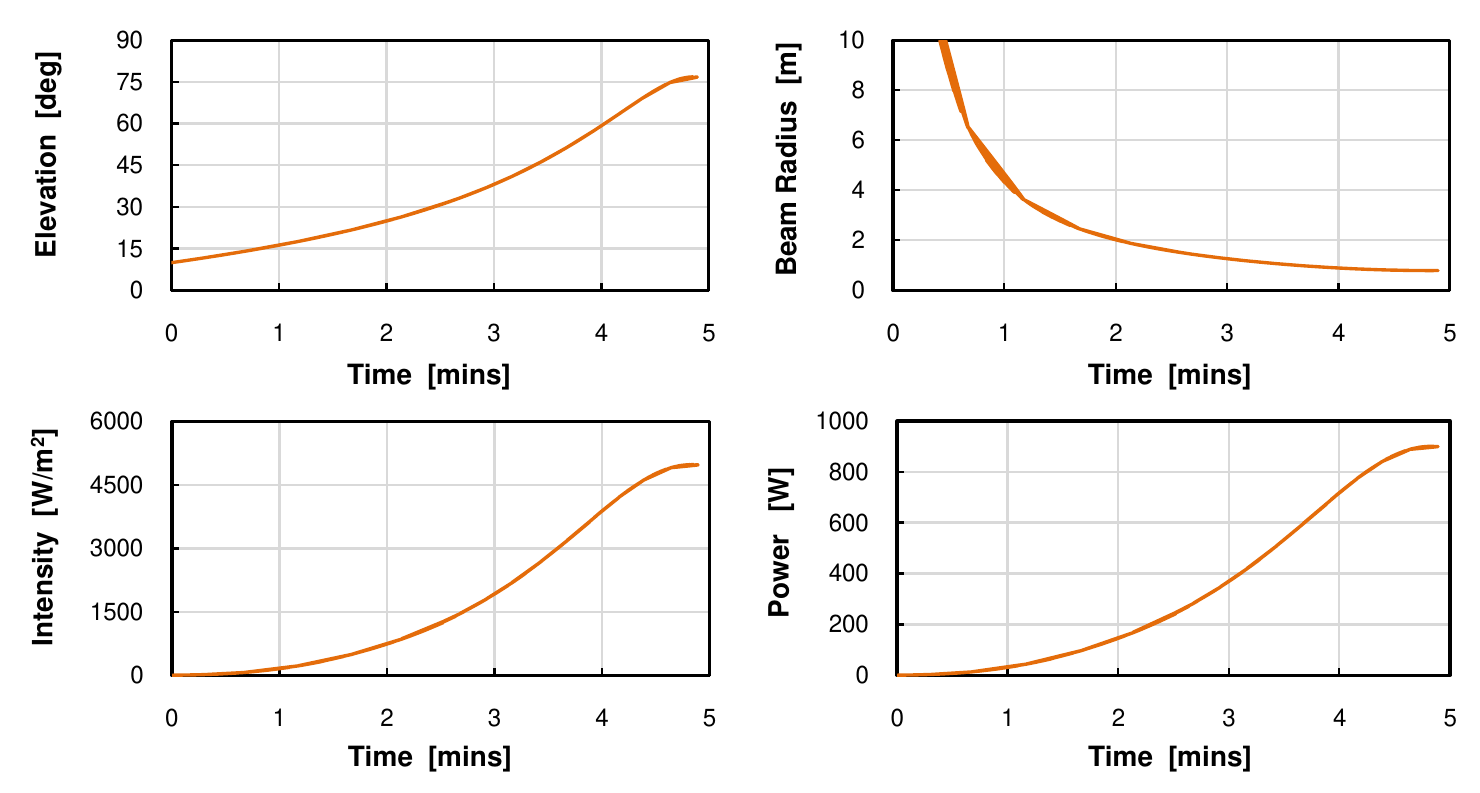}
\end{center}
\caption{The behavior of the beam as it tracks the Akari lens cap through a single near-overhead pass.}
\label{beamovertime}
\end{figure*}

\begin{figure*}
\begin{center}
\includegraphics*[width=14.4cm,angle=0]{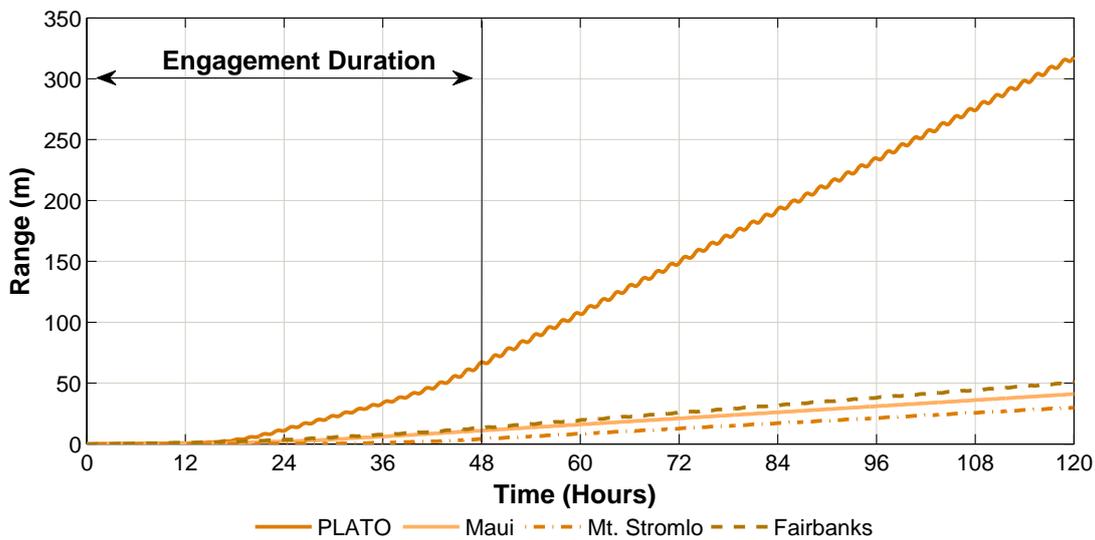}
\end{center}
\caption{Displacement of Akari lens cap from unperturbed orbit after 2 days of laser engagements, plotted for different system locations (for details see table \ref{successtable}).}
\label{capdisplacement}
\end{figure*}

Fig \ref{beamovertime} shows how the beam radius varies due to the changing beam path as the lens cap passes over the facility, with the engagement ending at the maximum elevation. The peak intensity (at the center of the beam) and resultant power on the target are a minimum at the lowest elevation and increase throughout the 5 minute pass.
The resulting displacements from 5\,kW laser engagements during the first half of each pass of the debris object over the laser during a 48 hour period (25 engagements in the case of PLATO) are compared in Fig \ref{capdisplacement} for four separate locations.
The in-track rate of displacement, or velocity difference, resulting from the illumination campaign is 82 m/day. Using the approach given in \citet{Levit2010}, an analysis of 70 days of TLEs for the lens cap showed that the orbital in-track error grows by an average of 178\,m/day, when propagated with a fitted numerical orbit propagator. . This method alone would not be sufficient to detect a maneuver on this object.However, a 10\,kW facility would generate 161 m/day which may well be detectable. 

For a conjunction of two objects with similar magnitude error to the Akari lens cap (and provided that one arranges the engagement geometry so as to increase the current predicted miss distance (e.g. by appropriately choosing between velocity vector and anti-velocity vector nudging)) such a system may be sufficient to significantly reduce the collision probability of a conjunction. With higher accuracy data based on any of (a) access to the U.S. Strategic Command unclassified SP catalog, (b) improved orbits obtained from tasked radar/optical tracking or (c)  TLE improvement scheme proposed  in Levit \& Marshall (2010), it is highly likely that the laser can provide more than sufficient $\Delta v$ to overwhelm the orbit/propagation errors, at least for objects of sufficiently high area to mass ratio. As an initial guide point, we will hereafter consider displacements of more than 200\,m/day as significant in that they are likely to overwhelm orbit errors associated with propagating high accuracy debris orbits.

\begin{figure}
\begin{center}
\includegraphics*[width=8.8cm,angle=0]{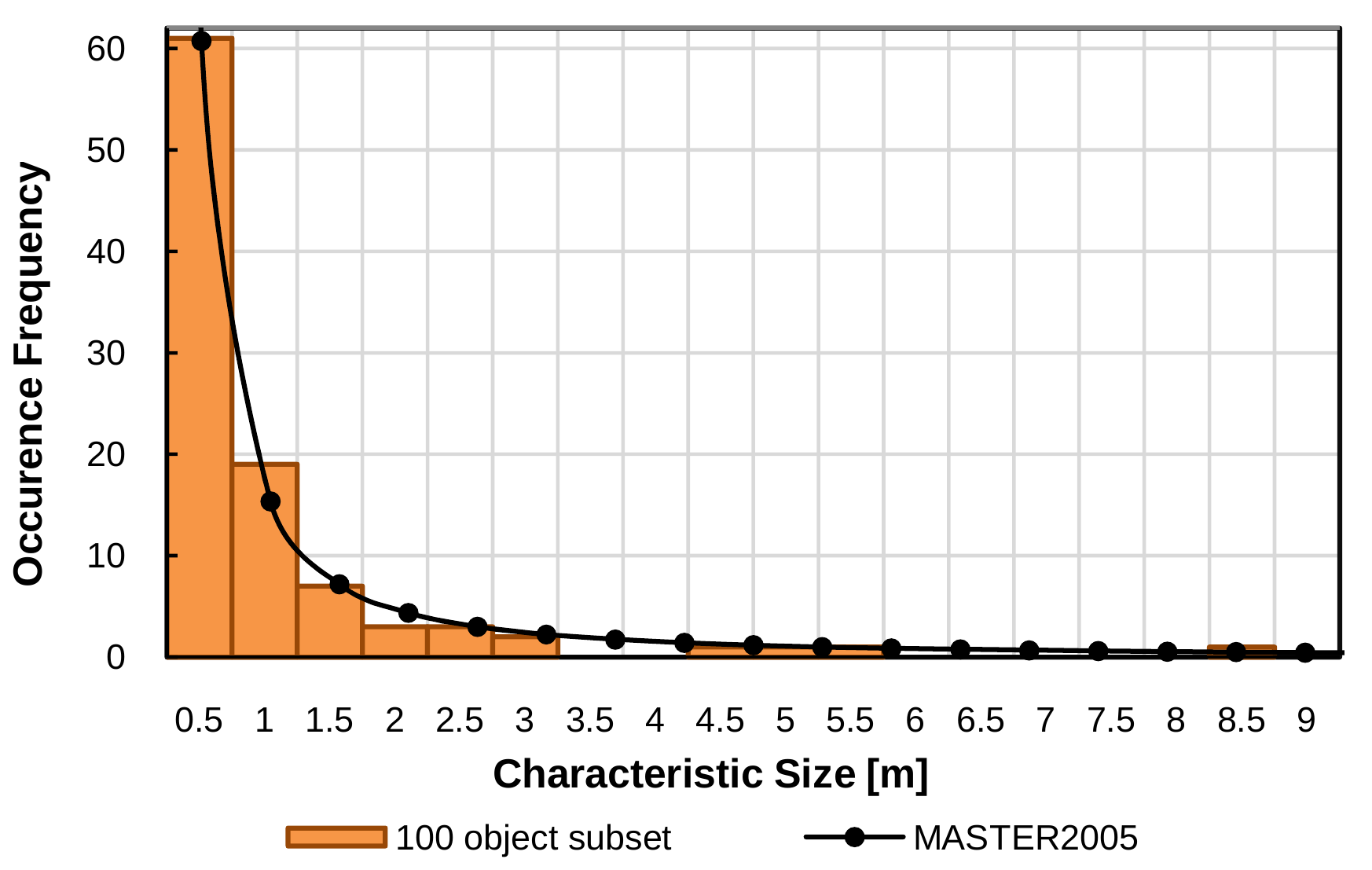}
\end{center}
\caption{Size distribution for 100 debris objects in sun-synchronous LEO, generated using MASTER2005's characteristic size distributions.}
\label{sizedistrisun100}
\end{figure}

We chose a random subset of 100 debris objects from the U.S. TLE catalog with inclinations between 97 and 102 degrees and orbit altitudes between 600 and 1100\,km. Our selection was limited to this number by the computational requirements of running these simulations. Characteristic sizes were assigned to these objects to give a representative size distribution, shown in Fig \ref{sizedistrisun100}.

\begin{figure}
\begin{center}
\includegraphics*[width=8.8cm,angle=0]{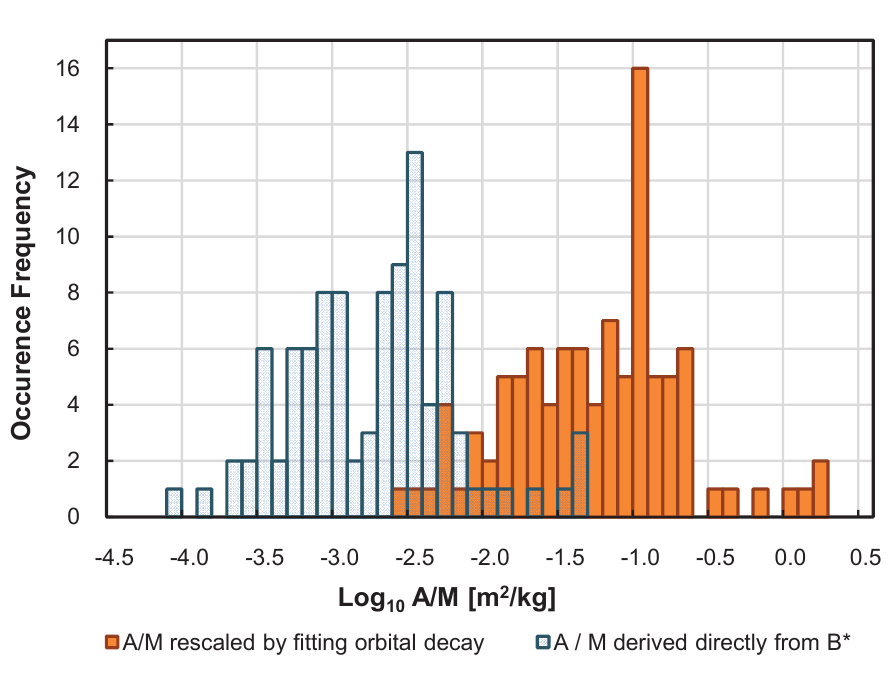}
\caption{Debris subset $A/M$ distribution, as inferred by a long term (120 day) statistical orbital decay assuming $C_d = 2.2$.}
\label{sizedistrisubset}
\end{center}
\end{figure}

The $A/M$ ratio of each object was determined (see Fig \ref{sizedistrisubset}) by rescaling the ballistic coefficient, allowing us to derive mass values for the set.

The mean $A/M$ after rescaling was $0.24$\,m$^2$/kg and the median was $0.11$\,m$^2$/kg. Two days was selected as a reasonable minimum conjunction warning lead time, during which the laser system could be employed. The laser was tasked with illuminating the target for the first half of each pass for 48 hours and the resultant displacement (from the unperturbed orbital position) was generated for the next five days.

As the size of the object increases beyond the beam width, the force on the object asymptotically approaches $F_{max} = C_r\times 1/c \times I_{max}\times\pi\times(1/2)\times w_{eff}^2$. There is therefore an upper limit on the mass of an object that can be sufficiently perturbed using laser applied photon pressure with any given system. This limit depends strongly on the geometry of the laser-target interaction, so we do not derive this limit analytically. To give an idea of this upper mass threshold, objects with masses greater than 100\,kg were all perturbed by less than 100\,m/day. As expected, photon pressure is generally not sufficient for maneuvering massive objects.

\begin{figure*}
\begin{center}
\includegraphics*[width=14.4cm,angle=0]{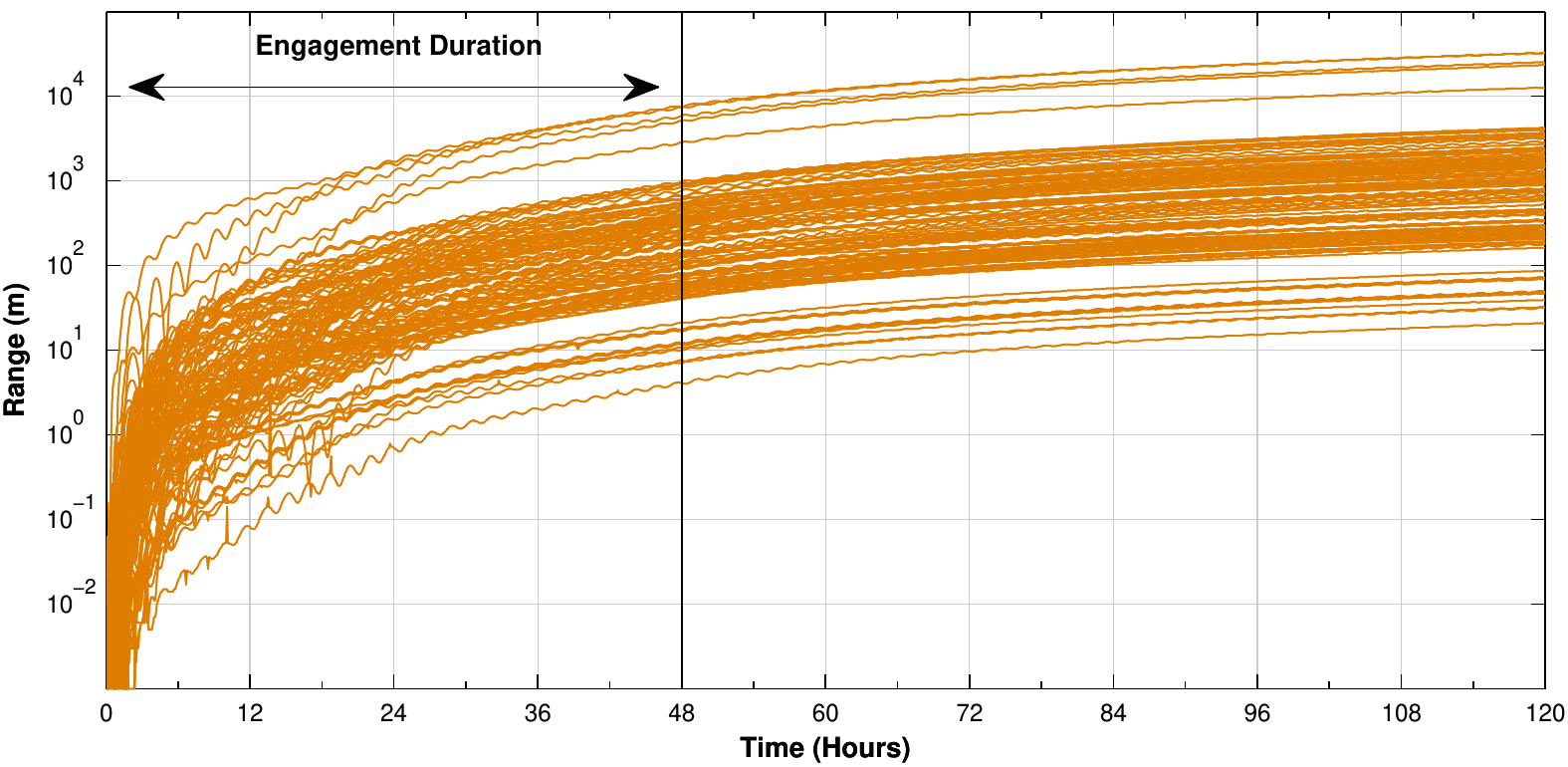}
\caption{Displacement from the unperturbed trajectory for 100 LEO debris fragment objects, each engaged by a 5 kW laser at PLATO at every opportunity for the first 48 hours. Displacements obtained using the 10 kW laser are approximately doubled.}
\label{100displaced}
\end{center}
\end{figure*}

For a single 5\,kW laser facility located at PLATO in Antarctica, the displacement from the unperturbed orbit for 100 objects is plotted in Fig \ref{100displaced}.
After a two day laser campaign it was found that 43 of 100 objects were diverging from their unperturbed orbit by more than 200 meters per day and 13 by more than 500 meters per day. For a 10\,kW laser, 56 objects where perturbed more than 200\,m and 34 more than 500\,m. A number of other ``success rates'', defined as the number of objects displaced by more than $x$ m/day, are shown in Table \ref{successtable}.
\begin{table*}[htbp]
\centering
\caption{Success Rates for 5 and 10 kW laser systems, compared for different sites. The Success Rates are defined as the number of objects displaced more than 50, 100, 200, 500 or 1000\,m/day.}
\begin{tabular}{lccccccrr}
\addlinespace
\toprule
\multicolumn{4}{c}{\textbf{Site Parameters}} & \multicolumn{5}{c}{\textbf{Success Rates (daily displacement)}} \\
\midrule
\textbf{Power / Location} & \textbf{Latitude} & \textbf{Longitude} & \textbf{Altitude} & \textbf{50\,m} & \textbf{100\,m} & \textbf{200\,m} & \textbf{500\,m} & \textbf{1000\,m}\\
{~5\,kW PLATO, Antarctica} & -80.37 & ~~77.35 & 4.09 km & 74 & 56 & 43 & \multicolumn{1}{c}{13} & \multicolumn{1}{c}{5} \\
{~5\,kW AMOS, Hawaii} & ~20.71 & -156.26 & 3.00 km & 30 & 13 & 5 & \multicolumn{1}{c}{4} & \multicolumn{1}{c}{2} \\
{~5\,kW Mt. Stromlo, Australia} & -35.32 & ~149.01 & 0.77 km & 11 & 4 & 4 & \multicolumn{1}{c}{3} & \multicolumn{1}{c}{0} \\
{~5\,kW Eielson AFB, Alaska} & ~64.85 & -148.46 & 0.50 km & 31 & 12 & 5 & \multicolumn{1}{c}{4} & \multicolumn{1}{c}{2} \\
\midrule
{10\,kW PLATO, Antarctica} & -80.37 & ~~77.35 & 4.09 km & 89 & 74 & 56 & \multicolumn{1}{c}{34} & \multicolumn{1}{c}{13} \\
{10\,kW AMOS, Hawaii} & ~20.71 & -156.26 & 3.00 km & 42 & 30 & 13 & \multicolumn{1}{c}{5} & \multicolumn{1}{c}{4} \\
{10\,kW Mt. Stromlo, Australia} & -35.32 & ~149.01 & 0.77 km & 29 & 12 & 4 & \multicolumn{1}{c}{4} & \multicolumn{1}{c}{3} \\
{10\,kW Eielson AFB, Alaska} & ~64.85 & -148.46 & 0.50 km & 48 & 31 & 12 & \multicolumn{1}{c}{4} & \multicolumn{1}{c}{4} \\
\bottomrule
\end{tabular}%
\label{successtable}%
\end{table*}%
Situating such a laser system in Antarctica may prove infeasible, so for comparison the simulation was run for the case of a single laser situated at the Air Force Maui Optical and Supercomputing site in Hawaii, at Mt. Stromlo in Australia and at a fictional location near Fairbanks, Alaska. Table \ref{successtable} shows the success rate of the system at these different locations for a 5\,kW and 10\,kW laser system.

Since the targets are all approximately sun synchronous the effectiveness of sites away from the polar region is greatly reduced, as expected. Mt. Stromlo and Maui show similar levels of performance. The additional atmospheric losses at Mt. Stromlo's lower altitude are offset by its higher latitude. Alaska performs better due to its higher latitude, but would benefit from being situated at higher altitude. The success rates shown in Table \ref{successtable} are meant to give a qualitative estimate of the campaign's effectiveness at avoiding collisions. The true effectiveness of a laser campaign is measured by re-evaluating the collision probability to determine whether it has decreased sufficiently to be confident of a miss. The collision probability is derived from the orbital covariance of the two objects, which was not available for this analysis. Therefore we do not perform a thorough collision probability analysis, but rather present the range displacements resulting from the simulated laser illumination campaign.

A 200\,m/day range displacement is equivalent to a $\Delta v$ impulse of about 0.08\,cm/s in the anti-velocity direction. Typical Envisat collision avoidance maneuvers have been of the order of a few cm/s, but were usually performed within a few hours of the conjunction epoch. Satellite operators want to minimize a maneuver's impact to the lifetime and mission schedule and therefore take the decision at the latest possible time to be sure that the maneuver is actually necessary. Additionally, for remote sensing satellites where lighting angles are important, maneuvers are often selected to quickly raise or lower the orbit to increase the radial miss distance, rather than rephrasing the satellite in True Anomaly, and/or they are combined with station-keeping maneuvers. For debris-debris collision avoidance using a laser this is not a concern and engagement campaigns may begin much earlier (i.e. two days before), letting small changes to the semi-major axis re-phase the target over longer periods.

\citet{Levit2010} suggest that batch least-squares fitting techniques can generate high accuracy orbital state vectors with errors that grow at about 100\,m/day. This error growth is of the same level as that provided by the high accuracy special perturbations catalog(s) maintained by the US Strategic Command \citep{Boers2000}. Given either of these sources, a range displacement of 200 m/day would dominate the growth of the object's error ellipse and would thus likely be sufficient for collision avoidance, but a full collision probability analysis is needed to confirm this. Additionally, data from initial engagements could reduce the size of the error ellipse, meaning that less range displacement (or, equivalently, less $\Delta v$) will be required to reduce the collision probability.

\section{Discussion on Next Steps and Implications}
\subsection{Further Research}
Immediate follow up work should focus on reducing the uncertainty of modeling assumptions to improve the statistics presented here. Near-term improvements should include the following:
\begin{enumerate}
\item Test the effect of this scheme in long-term evolutionary models, such as the NASA LEGEND model \citep{Liou2004}. By considering the long term consequences of shielding the ``high impact'' population (objects of both high collision cross-section and large mass) from the type of objects for which photon pressure is effective we could determine how many objects would need to be shielded to halt the cascading growth of debris in low Earth orbit. This would provide a better prediction of the long term effectiveness of the system. 
\item Radar Cross Section (RCS) data might be used to determine the characteristic sizes for individual debris objects, instead of - as we have done - using randomly assigned sizes that match the observed distribution. This would allow simulations using more accurate object areas and masses. There is some uncertainty in the accuracy of RCS measurements, and further research and analysis should be conducted before adopting this approach.
\item The simulations should be run for a much larger set of objects, and in a wider range of orbital regimes, to allow useful statistics to be generated and a metric devised to identify the class of objects for which the system is truly effective.
\item Error covariances should be generated for each simulated object's orbit. This would allow us to estimate the change in collision probability resulting from consecutive engagements, a far more useful measure of the systems capability than the simple range displacement.
\item A systematic parameter optimization study needs to be done to identify the best combination of laser power versus number of facilities, the optimal locations for these facilities, the most advantageous engagement strategy and the ideal combination of laser and optics.
\item Spin assessment. Since the illuminations can provide torques to the debris objects being illuminated, it is prudent to research, in detail, the effects that this could have. For example, changing the spin rates could alter the drag coefficient and make it harder to predict the orbit position of that debris object. It would also effect the decay lifetime, potentially making it longer. Finally, it could reduce the object's radar cross-section. None of these issues seem on first analysis to be a significant challenge to the system's overall utility but they demand detailed consideration.
\item Finally, the policy implications need consideration. These include the problems of debris ownership (and potential need for transfer of that ownership) and associated liability of maneuvering a piece of debris. There are also potential security concerns for the system which may demand solutions similar to laser de-confliction, as practiced by the ILRS \citep{Pearlman2002}.
\end{enumerate}

\subsection{Technology Demonstration}
Following the aforementioned further research and a comprehensive engineering and costing analysis, a technical demonstration would be the logical next step. This could most easily be accomplished by integrating a continuous wave fiber laser (and adaptive optics if necessary) into an existing fast slewing optical telescope and demonstrating the acquisition, tracking and orbit modification of a known piece of debris (a US-owned rocket shroud for example). The thermal, mechanical and optical implications of continuous 5\,kW IR laser operations would need to be addressed via engineering simulation first, and probably verified in actual tests. Eventual candidates for a demonstration include the EOS Mt. Stromlo facility and the Advanced Electro-Optical System at AMOS. AEOS has demonstrated large-aperture debris tracking with the ~180W HI-CLASS ladar system \citep{Kovacs2001}. EOS is routinely performing laser tracking of LEO debris objects smaller than 10\,cm in size from this facility\citep{Greene2002}. The EOS facility would probably require the fewest modifications to incorporate a higher power CW fiber laser for a technology demonstration. Since the 5\,kW laser costs \$0.8M, we speculate that the direct cost of adapting such a system would be of order \$1-2M. In addition, it may be possible to perform a near-zero cost demonstration using existing capabilities such as those of the Starfire Optical Range at Kirtland AFB. It should be noted that the authors know of no relevant system that already has adaptive optics capable of fast slew compensated beam delivery to LEO.

Having demonstrated the method on an actual piece of debris, a fully operational system could be designed and located at an optimal site, or appended to a suitable existing facility. Preliminary discussions with manufacturers suggest that the capital cost of the laser and primary beam director would be around \$3-6M. The cost of the necessary primary adaptive optics and tracking systems (including secondary lasers and tracking optics) are less clear at this stage since there are a number of ways that a working solution could be engineered. Further engineering analysis is necessary before accurate overall system costs can be estimated. There is advantage to making the system an international collaboration in order to share cost, to ease certain legal obstacles to engaging space objects with varied ownership and to reduce the likelihood of the facility being viewed negatively from a security stand point. This system would coincidentally complete many of the steps (both technical and political) necessary to implement an ORION-class laser system to de-orbit debris, potentially clearing LEO of small debris in just a few years \citep{Phipps1996}, if it was deemed useful to do that in addition. A key component for the proposal herein would also be an operational all-on-all conjunction analysis system, the cost of which is also uncertain but likely to be small compared to the other system costs to operate (and which would also benefit from including multiple international datasets).

\subsection{Potential Implications for the Kessler Syndrome}
\citet{Liou2009} have identified the type of ``high impact'' large mass, large area objects that will drive the growth of the LEO debris population from their catastrophic collisions. In the LEO sun synchronous region the high impact debris mass is approximately evenly divided between large spacecraft and upper rocket bodies \citep{Liou2011}. ESA routinely monitors all conjunctions with objects predicted to pass through a threat volume of 10\,km$ \times $25\,km$ \times $10\,km around its Envisat, ERS-2 and Cryosat-2 satellites using their Collision Risk Assessment tool (CRASS). These satellites are operational and maneuverable, but their orbit and mass and area profiles' make them analogous to Liou and Johnson's high impact objects. We therefore use these satellites as a proxy for the high impact population.

75\% of conjunctions with Envisat's threat volume involve debris (i.e. not mission related objects, rocket bodies or other active spacecraft). Significantly, 61\% of all Envisat conjunctions involve debris resulting directly from either the Fengyun 1-C ASAT test or from the Iridium 33/Cosmos 2251 collision. For ERS-2 and Cryosat-2 (at a lower altitude) these figures are similar \citep{Flohrer2009}. It is clear that debris resulting primarily from collision and explosion fragments is most likely to be involved in collisions with large objects in the LEO polar region.

These statistics suggest that it may be possible to shield high impact objects from a significant proportion of catastrophic collisions with less massive debris such as fragments by using a ground based medium power laser. If 75\% of conjunctions with high impact objects involve debris (as suggested by Envisat conjunctions) and our analysis of 100 random debris objects suggest that 43\% can be significantly ($>$200\,m/day) perturbed using our baseline 5\,kW system, then it may be possible to prevent a third of all conjunctions involving the high impact population. Increasing the laser power to 10\,kW would raise this figure to 42\%.

Additionally, LEGEND simulations have shown that catastrophic collisions involving intacts (spacecraft and rocket bodies) and fragments are slightly more likely than collisions involving only intacts \citep{Liou2011}. Using these collision statistics, and assuming 200\,m/day is sufficient to insure a clear miss, we see that a single 5\,kW system could prevent nearly half of all catastrophic collisions involving debris fragments, and about 28\% of all collisions, including intact-intact collisions. Obviously an intact-intact collision is a bigger debris source than an intact-fragment of fragment-fragment since it involves two massive objects. Further LEGEND modeling would be able to quantify the degree to which the scheme reduces debris sources.

Of course one is not limited to shielding one object. We posit that it may be possible to use laser photon pressure as a substitute for active debris removal, provided a sufficient number of high impact objects can be continually shielded to make the two approaches statistically similar. Indeed, the routine active removal of 5 large debris objects per year is predicted to prevent 4 intact-intact, and 5 intact-fragment catastrophic collisions over the next 200 years \citep{Lious2011}. With an effective all-on-all conjunction analysis system to prioritize engagements and considering that every engagement reduces the target's orbital covariance (thereby halting unnecessary engagement campaigns) it is plausible that far more objects may be shielded than are required to make the two approaches equivalent in terms of preventing the number of catastrophic collisions (a LEGEND simulation may confirm this). 

For a facility on the Antarctic plateau the laser would be tasked to an individual object for an average of 103 minutes per day. The laser can only track one target at a time, but average pass times suggest that it is possible to optimize a facility to engage $\sim$10 objects per day. The Envisat conjunction analysis statistics suggest around 10 high risk (above 1:10,000) events per high impact object, per year \citep{Flohrer2009}. If improved accuracy catalogs or tracking data become available then it is feasible that the system could engage thousands of (non-high impact) objects per year, or conversely that up to hundreds of high impact objects could be shielded by one facility per year. This is an order of magnitude more objects than one needs to remove in order to stabilize the growth \citep{Liou2009}. Preventing collisions on such a large scale would therefore likely reduce the rate of debris generation such that the rate of debris reentry dominates and the Kessler syndrome is reversed at low enough altitudes. Continued operation over a period similar to the decay timescale from the orbital regions in question (typically decades) could thus reverse the problem. Additionally, scaling such a system (eg. multiple facilities) on the ground would be low cost (relative to space missions) and can be done with currently mature technology, making it a good near term solution. Further, if the current analysis proves optimistic, raising the power to 10\,kW and having 3-4 such facilities would increase the number of conjunctions that it is possible to mitigate by a further order of magnitude, and also would raise the maximum mass and reduce the minimum $A/M$ threshold for the system.

\subsection{Additional Applications}
The described system has a number of alternative uses, which may further improve the value proposition.

Firstly, orbit tracks are a byproduct of target acquisition that can be used for orbit determination. Correlating these tracks would allow the generation of a very high accuracy catalog, similar to that being produced by the EOS facility at Mt. Stromlo. The return signal from laser illumination will potentially provide data for accurate estimation of debris albedo and, if the object is large enough to be resolved, size, attitude and spin state; thus helping space situation awareness more generally.

Secondly, the concept of shielding high impact debris objects can be applied to protecting active satellites. The laser system could begin engaging the debris object following a high risk debris-satellite conjunction alert. The initial engagements would provide additional orbit information that may reduce the risk to an acceptable level. Continued engagement would perturb the debris orbit, potentially saving propellant by avoiding the need for a satellite maneuver. This could even be provided as a commercial service to satellite operators wishing to extend operation lifetimes by saving propellant.

Lastly the laser system may also prove useful for making small propellant-less maneuvers of satellites, including those without propulsion, provided the satellite is sufficiently thermally protected to endure 5-minute periods of illumination with a few times the solar constant. This could be used to, for example, enable formation-flying clusters of small satellites, or perform small station-keeping maneuvers. Being able to extend smallsat lifetimes without launching to higher altitudes or being able to gradually re-phase a satellite in True Anomaly may also have commercial applications.

\section{Conclusion}
It is clear that the actual implementation of a laser debris-debris collision avoidance system requires further study. Assumptions regarding the debris objects’ properties need refinement and a detailed engineering analysis is necessary before a technology demonstration can be considered. However, this early stage feasibility analysis suggests that a near-polar facility with a 5\,kW laser directed through a 1.5\,m fast slewing telescope with adaptive optics can provide sufficient photon pressure on many low-Earth sun-synchronous debris fragments to substantially perturb their orbits over a few days. Additionally, the target acquisition and tracking process provides data to reduce the uncertainties of predicted conjunctions. The laser need only engage a given target until the risk has been reduced to an acceptable level through a combination of reduced orbital covariance and actual photon pressure perturbations. Our simulation results suggest that such a system would be able to prevent a significant proportion of debris-debris conjunctions.

Simulation of the long term effect of the system on the debris population is necessary to confirm our suspicion that it can effectively reverse the Kessler syndrome at a lower cost relative to active debris removal (although quite complementary to it). The scheme requires launching nothing into space - except photons - and requires no on-orbit interaction - except photon pressure. It is thus less likely to create additional debris risk in comparison to most debris removal schemes. Eventually the concept may lead to an operational international system for shielding satellites and large debris objects from a majority of collisions as well as providing high accuracy debris tracking data and propellant-less station keeping for smallsats.

\section{Acknowledgments}
We would like to thank the following individuals for useful conversations and contributions: Luciano Anselmo, John Barentine, Tim Flohrer, Richard L. Garwin, R\"udiger Jehn, Kevin Parkin, Brian Weeden, S. Pete Worden and the anonymous reviewers of this paper.

\end{document}